\documentclass[aps,pre,reprint,groupedaddress,amsmath,amsfonts,amssymb,letterpaper]{revtex4-2}
\usepackage{graphicx}
\usepackage{subfigure}
\usepackage{tikz}
\usepackage{mathtools}
\usepackage{appendix}

\usepackage[subtle]{savetrees}
\usepackage{amsmath,amsfonts,amssymb}
\usepackage{graphicx}
\usepackage{subfigure}
\usepackage{mathtools}
\usepackage{makecell}
\usepackage{mathrsfs}
\usepackage{enumitem}
\usepackage{hyperref}
\hypersetup{colorlinks=true,linkcolor=blue,urlcolor=blue,citecolor=blue}

\newcommand{\sO}{\mbox{\footnotesize{$(\Omega)$}}}
\newcommand{\saO}{\mbox{\footnotesize{$(\alpha\Omega)$}}}
\DeclareMathOperator{\lcm}{lcm}

\usetikzlibrary{matrix,decorations.text,decorations.pathmorphing,decorations.markings,arrows,calc,shapes.geometric,patterns,shadows,intersections}
\begin{document}
\title{Controlling the direction of steady electric fields in liquid using\\non-antiperiodic potentials}
	
\author{Aref Hashemi}
\email[Email: ]{aref@cims.nyu.edu}
\affiliation{Courant Institute, New York University, New York, NY, United States}
\author{Mehrdad Tahernia}
\affiliation{Independent Researcher}
\author{William D. Ristenpart}
\email[Email: ]{wdristenpart@ucdavis.edu}
\author{Gregory H. Miller}
\email[Email: ]{grgmiller@ucdavis.edu}
\affiliation{Department of Chemical Engineering, University of California Davis, Davis, CA, United States}

\begin{abstract}
When applying an oscillatory electric potential to an electrolyte solution, it is commonly assumed that the choice of which electrode is grounded or powered does not matter because the time-average of the electric potential is zero. Recent theoretical, numerical, and experimental work, however, has established that certain types of multimodal oscillatory potentials that are ``non-antiperodic'' can induce a net steady field toward either the grounded or powered electrode [Hashemi et al., Phys. Rev. E \textbf{105}, 065001 (2022)]. Here, we elaborate on the nature of these steady fields through numerical and theoretical analyses of the asymmetric rectified electric field (AREF) that occurs in electrolytes where the cations and anions have different mobilities. We demonstrate that AREFs induced by a non-antiperiodic electric potential, e.g., by a two-mode waveform with modes at $2$ and $3$ $\mathrm{Hz}$, invariably yields a steady field that is spatially dissymmetric between two parallel electrodes, such that swapping which electrode is powered changes the direction of the field. Additionally, using a perturbation expansion, we demonstrate that the dissymmetric AREF occurs due to odd nonlinear orders of the applied potential. We further generalize the theory by demonstrating that the dissymmetric field occurs for all classes of zero-time-average (no dc bias) periodic potentials, including triangular and rectangular pulses, and we discuss how these steady fields can tremendously change the interpretation, design, and applications of electrochemical and electrokinetic systems.
\end{abstract}
	
\maketitle

\section{\uppercase{Introduction}}
Application of ac electric potentials to liquids is a ubiquitous element of electrokinetic systems, including induced-charge-electrokinetics (ICEK) \cite{BazantPRL2004,Squires2004}, ac electroosmosis (ACEO) \cite{Ramos1998,Ramos1999,Ajdari2000,Studer2004}, and electrohydrodynamic (EHD) manipulation of colloids \cite{Trau1996,Prieve2010,Ma2012,Dutcher2013}. Over the last few decades, a great body of research has focused on evaluating the dynamic response of liquids to ac polarization, in order to find the induced electric field and ion concentrations within the liquid \cite{BazantPRE2004,Bazantreview2009,Ramos2016}. However, ion-containing liquids respond to ac polarizations in intricate ways, especially when the dissolved ions have unequal mobilities. In particular, recent studies have established the existence of an induced, long-range, steady field in liquids, referred to as asymmetric rectified electric field (AREF) \cite{Aref2018,Aref2019,Aref2020SM,Balu2021}. A perfectly sinusoidal potential induces an electric field with a nonzero time-average, a zero-frequency component, as a direct result of the nonlinear effects and ionic mobility mismatch. AREF was shown to provide qualitative explanations for several long-standing questions in electrokinetics and significantly change the interpretation of experimental observations \cite{Aref2018,Bukosky2019,Aref2020}.

For a sinusoidal applied potential of amplitude $\phi_0$ and angular frequency $\omega$, the one-dimensional AREF between parallel electrodes is antisymmetric with respect to the midplane \cite{Aref2018}. Depending on the applied frequency, electrolyte type, and electrode spacing, AREF may change sign several times within the liquid \cite{Aref2019}. However, it remains identically zero at the midplane and at the electrodes. Such an antisymmetric shape indicates that the AREF does not change upon swapping the powered and the grounded electrodes, or introducing any time or phase lag to the applied potential.
	
However, the aforementioned characteristics of AREF do not necessarily hold for other classes of zero-time-average (no dc bias) periodic potentials. In fact, a recent study by Hashemi et al. \cite{Aref2022PRE} shows that oscillatory potentials with a certain time symmetry break can induce AREFs that are dissymmetric (as different from antisymmetric) in space. Such behavior is a reminiscent of so called ``temporal ratchets,'' a well-known phenomenon in the context of point particles and optical and quantum lattice systems \cite{Flach2000,Denisov2002,Ustinov2004,Denisov2014}. Here, we provide an extensive numerical and theoretical analysis of the ratchet AREF, its origin, and its important implications to electrokinetics. 

\begin{figure}[t]
\begin{center}
\centering
\setlength{\belowcaptionskip}{-20pt}
\includegraphics{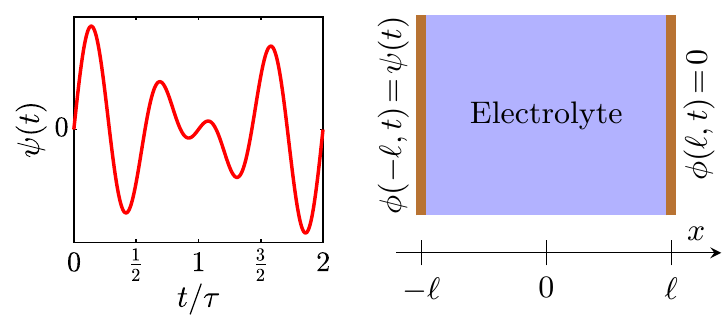}
\caption{Schematic diagram of the problem. An electrolyte confined between parallel, planar, electrodes, separated by a gap $2\ell$, and powered by a multimodal potential $\psi(t)$ with period $2\tau$.}
\label{Fig:schematic}
\end{center}
\end{figure}

\section{\uppercase{Problem Statement}}
Consider a dilute binary $1$--$1$ electrolyte confined by two parallel, planar, electrodes spaced by a gap $2\ell$ (\autoref{Fig:schematic}). A two-mode potential $\psi(t)=\phi_0\left[\sin(\omega t)+\sin(\alpha\omega t)\right]$, with $\alpha$ a rational number, is applied on the electrodes as
\begin{equation}
  \phi(-\ell,t)=\psi(t),\quad \phi(\ell,t)=0.
  \label{Eq:BCphi}
\end{equation}

The starting point in theory to investigate the dynamics of such a system is the Poisson--Nernst--Planck (PNP) model. The Poisson equation relates the free charge density to the electric field gradient,
\begin{equation}
  -\varepsilon\frac{\partial^2\phi}{{\partial x}^2}=\rho=e(n_+-n_-),
  \label{Eq:Poisson}
\end{equation}
while the transport of ions is governed by the Nernst--Planck equations,
\begin{equation}
  \frac{\partial n_{\pm}}{\partial t}=D_{\pm}\frac{\partial^2n_{\pm}}{{\partial x}^2}\pm \frac{D_{\pm}}{\phi_T}\frac{\partial}{\partial x}\Big(n_{\pm}\frac{\partial\phi}{\partial x}\Big).
\label{Eq:SCE}
\end{equation}
Here the symbols denote permittivity of the electrolyte, $\varepsilon$; electric potential, $\phi$; free charge number density, $\rho$; charge of a proton, $e$; thermal potential, $\phi_T$; ion number concentration, $n_{\pm}$; diffusivity, $D_{\pm}$; location with respect to the midplane, $x$; and time, $t$.
	
Initially, the ions are uniformly distributed $n_{\pm}(x,0)=n^{\infty}$ (the bulk electrolyte concentration), and the electric potential is zero everywhere $\phi(x,0)=0$. Note that for simplicity, we neglect the intrinsic zeta potential of the electrodes. Finally, at $x=\pm\ell$ (i.e., the electrodes), we set the flux of ions equal to zero (i.e., no electrochemistry).
	
\section{\uppercase{Numerical Results \& Discussion}}
\begin{figure}[t]
\begin{center}
\centering
\setlength{\belowcaptionskip}{-20pt}
\includegraphics{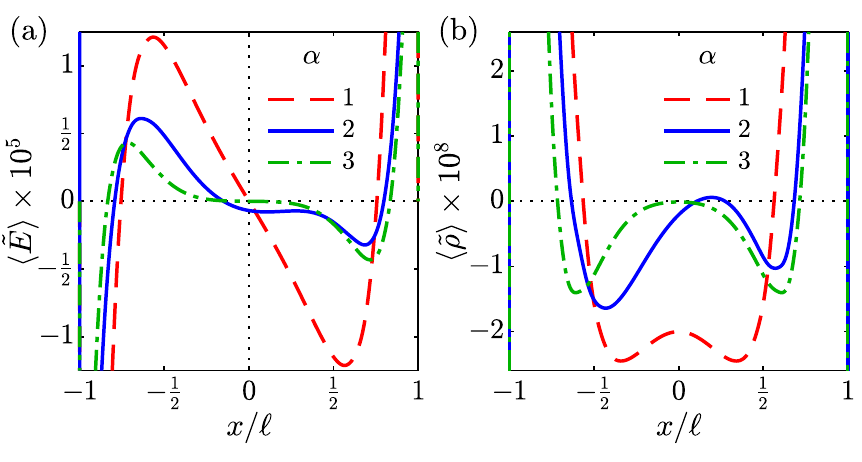}
\caption{Representative numerical solutions to the AREF $\langle \tilde{E}\rangle=\langle E\rangle /(\kappa\phi_T)$, (a), and time-average free charge density $\langle \tilde{\rho}\rangle=\langle \rho\rangle/n^{\infty}$, (b), for two-mode applied potentials ($\phi(t,-\ell)=\phi_0\left[\sin(\omega t)+\sin(\alpha\omega t)\right]$, $\phi(\ell,t)=0$) in the bulk electrolyte. Parameters: $\phi_0=10\phi_T$, $f=\omega/(2\pi)=50$ $\mathrm{Hz}$, $2\ell=20$ $\mu\mathrm{m}$, $D_+=10^{-9}$ $\mathrm{m^2/s}$, $D_-/D_+=2$, $c^{\infty}=1$ $\mathrm{mM}$.}
\label{Fig:AREF_two_mode_micronScale}
\end{center}
\end{figure}

The system of equations is solved numerically following the algorithm reported by Hashemi et al. \cite{Aref2018}. We focus primarily on the time-average of the harmonic solutions defined by
\begin{equation}
\langle \chi\rangle=\frac{1}{2\tau}\int_{t}^{t+2\tau}\chi dt,\quad 2\tau=\frac{1}{\gcd(1,\alpha)}\frac{2\pi}{\omega},
\end{equation}
where $2\tau$ is the period of the applied potential (or that of the harmonic solution), and $\gcd(1,\alpha)$ is the greatest common divisor of $1$ and $\alpha$ \footnote{For two rational numbers $a_1$ and $a_2$, $\gcd(a_1,a_2)$ can be computed by $\gcd(N_1,N_2)/\lcm(D_1,D_2),$ where $N_{1,2}$ and $D_{1,2}$ are, respectively, the integer numerators and denominators of $a_{1,2}$ written as fractions, and $\lcm$ denotes the least common multiple operator.}. Representative solutions to the AREF (time-average electric field) in the bulk electrolyte (i.e., several Debye layer lengths away from the electrodes) are provided in \autoref{Fig:AREF_two_mode_micronScale}(a). When $\alpha=1$, the applied potential is a single-mode sinusoid which yields the antisymmetric AREF (\autoref{Fig:AREF_two_mode_micronScale}(a), dashed red curve). The case of $\alpha=2$ reveals a surprising phenomena: the shape of the AREF becomes dissymmetric with a nonzero value even at the midplane (\autoref{Fig:AREF_two_mode_micronScale}(a), solid blue curve). Further complicating matters, for $\alpha=3$, the AREF is again perfectly antisymmetric (\autoref{Fig:AREF_two_mode_micronScale}(a), dash-dotted green curve). Therefore, it appears that depending on $\alpha$, the induced AREF can be either antisymmetric with a zero value at the midplane or dissymmetric. The corresponding spatial distributions of the time-average free charge density $\langle\rho\rangle$ are illustrated in \autoref{Fig:AREF_two_mode_micronScale}(b) for different $\alpha$ values. Consistent with the AREF distributions in \autoref{Fig:AREF_two_mode_micronScale}(a), $\langle\rho\rangle$ is spatially even for $\alpha=1$ and $3$, but takes a dissymmetric shape for $\alpha=2$.

\begin{figure}[t]
\begin{center}
\centering
\setlength{\belowcaptionskip}{-20pt}
\includegraphics{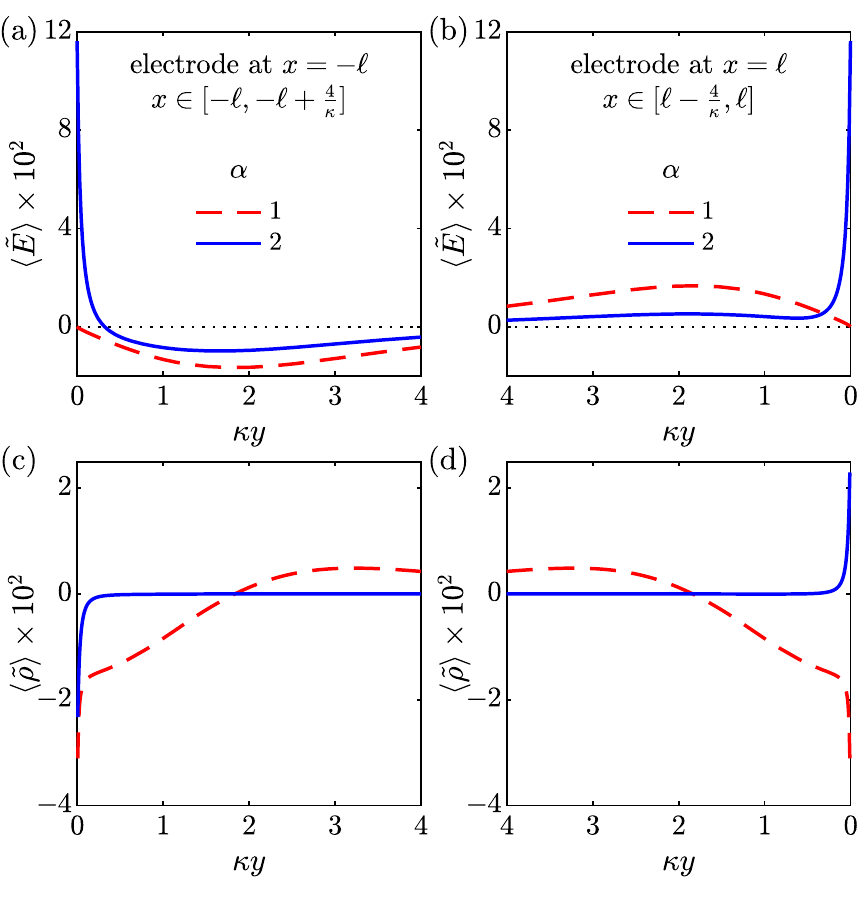}
\caption{Representative numerical solutions to the AREF $\langle \tilde{E}\rangle=\langle E\rangle /(\kappa\phi_T)$, (a, b), and time-average free charge density $\langle \tilde{\rho}\rangle=\langle \rho\rangle/n^{\infty}$, (c, d), for two-mode applied potentials ($\phi(t,-\ell)=\phi_0\left[\sin(\omega t)+\sin(\alpha\omega t)\right]$, $\phi(\ell,t)=0$) at the Debye scale. For visualization purposes, the $\langle \tilde{\rho}\rangle$ data for $\alpha=2$ in (c, d) are divided by $100$. The spatial variable $y$ denotes the distance from the corresponding electrode. Parameters: $\phi_0=10\phi_T$, $f=\omega/(2\pi)=50$ $\mathrm{Hz}$, $2\ell=20$ $\mu\mathrm{m}$, $D_+=10^{-9}$ $\mathrm{m^2/s}$, $D_-/D_+=2$, $c^{\infty}=1$ $\mathrm{mM}$.}
\label{Fig:AREF_two_mode_DebyeScale}
\end{center}
\end{figure}
	
The behavior becomes more complicated at the Debye scale (i.e., up a few Debye lengths away from the electrodes). \autoref{Fig:AREF_two_mode_DebyeScale}(a) and (b) show the AREF within $4$ Debye lengths away from the electrodes for $\alpha=1$ and $2$. When $\alpha=1$ (i.e., a single-mode sinusoidal potential), AREF is zero at the electrodes, which is a direct result of the antisymmetric shape of the AREF and the total charge neutrality. The former can be clarified by a parity analysis of the second-order perturbation solution (in terms of the applied potential) to the problem (cf. \autoref{sec:perturbation}). The total charge neutrality on the other hand enforces the AREF at one electrode to be equal to that on the other electrode, that is $\langle E\rangle_{-\ell}=\langle E\rangle_{\ell}=K$ for some constant $K$. But, for AREF to be antisymmetric $K$ has to be zero.

When $\alpha=2$, an astonishingly large AREF is induced on the electrodes ($\approx4$ orders of magnitude larger than the AREF in the bulk electrolyte). We note, however, that the total charge neutrality still holds. The mere observation of a nonzero AREF at the electrodes for $\alpha=2$ is consistent with the dissymmetric shape of AREF in the bulk electrolyte: the integral of the AREF over the entire domain has to be zero, i.e., $\int_{-\ell}^{\ell}\langle E\rangle dx=\langle \phi\rangle_{-\ell}-\langle \phi\rangle_{\ell}=0$. In other words, the nonzero AREF at the electrodes and the dissymmetric shape of the AREF in the bulk electrolyte are interrelated. A qualitatively consistent behavior is observed for the distribution of $\rho$ (\autoref{Fig:AREF_two_mode_DebyeScale}(c) and (d)). The induced $\langle\rho\rangle$ on the two electrodes are the same for $\alpha=1$. However, when $\alpha=2$, there is a sign flip in the time-average free charge densities induced at the two electrodes ($\langle \rho\rangle_{-\ell}=-\langle \rho\rangle_{\ell}$).
	
\begin{figure}[t]
\begin{center}
\centering
\setlength{\belowcaptionskip}{-20pt}
\includegraphics{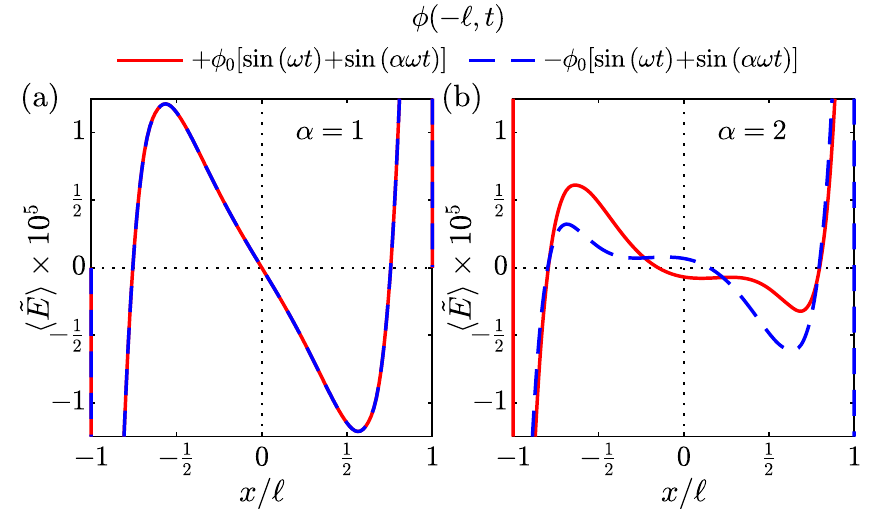}
\caption{Flipping the sign of the applied potential at $x=-\ell$ for $\alpha=1$ (a), and $2$ (b). Parameters: $\phi_0=10\phi_T$, $f=\omega/(2\pi)=50$ $\mathrm{Hz}$, $2\ell=20$ $\mu\mathrm{m}$, $D_+=10^{-9}$ $\mathrm{m^2/s}$, $D_-/D_+=2$, $c^{\infty}=1$ $\mathrm{mM}$.}
\label{Fig:AREF_reversePotential}
\end{center}
\end{figure}

We now ask what happens if we flip the sign of the applied potential ($-\psi(t)$ instead of $\psi(t)$). For $\alpha=1$ (antisymmetric AREF), the curves of the induced AREF by $\psi(t)$ and $-\psi(t)$ potentials are superimposed (\autoref{Fig:AREF_reversePotential}(a)). However, flipping the sign of the potential when $\alpha=2$ (dissymmetric AREF) yields a mirrored version of the AREF with respect to the midplane (\autoref{Fig:AREF_reversePotential}(b)). It is worth mentioning that the sum of the solid red ($\psi(t)$) and dashed blue ($-\psi(t)$) curves in \autoref{Fig:AREF_reversePotential}(b) is antisymmetric and zero at the midplane. In other words, the dissymmetric components of the AREFs due to $\psi(t)$ and $-\psi(t)$ potentials cancel each other.
	
We provide an explanation for this numerical observation using symmetry arguments. Note that the field-induced ion motion depends only on the potential gradient (not the potential itself). Therefore, one can show that flipping the sign of a periodic, time-varying, potential $\psi(t)$ at $x=-\ell$ is equivalent to swapping the powered and grounded electrodes (by adding the potential $\psi(t)$ to the both electrodes). In other words, $\{\phi(-\ell,t)=-\psi(t),\phi(\ell,t)=0\}\equiv\{\phi(-\ell,t)=0,\phi(\ell,t)=\psi(t)\}$. Now, a simple change of variable $x\to-x$ clarifies that if the potential $\psi(t)$ yields the electric field $E(x,t)$, the potential $-\psi(t)$ would yield the mirrored version, $-E(-x,t)$ (cf. \autoref{Fig:AREF_reversePotential}(b)).
	
Focusing on the midplane ($x=0$), one can write that the functional $E(0,t)=\epsilon(t)=f(\psi,t)$ is odd in $\psi$. Therefore, if $\epsilon(t)$ is the induced electric field at the midplane due to the potential $\psi(t)$, $-\epsilon(t)$ would be that due to the potential $-\psi(t)$. Now consider antiperiodic potentials, i.e., $\psi(t+\tau)=-\psi(t)$. We prove that $\langle\epsilon\rangle$ (i.e., AREF at the midplane) has to be zero for antiperiodic potentials:
\begin{equation}
\epsilon(t+\tau)=f(\psi,t+\tau)=f(-\psi,t)=-\epsilon(t).
\end{equation}
Therefore, $\epsilon(t+\tau)=-\epsilon(t)$, which upon taking a time-average yields $\langle\epsilon\rangle=-\langle\epsilon\rangle$, indicating $\langle\epsilon\rangle=0$. It is worth mentioning that the above argument is general and holds for any antiperiodic potential $\psi(t)$. It appears that for such potentials the zero-frequency components of the induced electric field cancel each other at the midplane, yielding an antisymmetric AREF. However, they do not necessarily cancel out when the excitation is non-antiperiodic.
	
\begin{figure}[t]
\begin{center}
\centering
\setlength{\belowcaptionskip}{-20pt}
\includegraphics{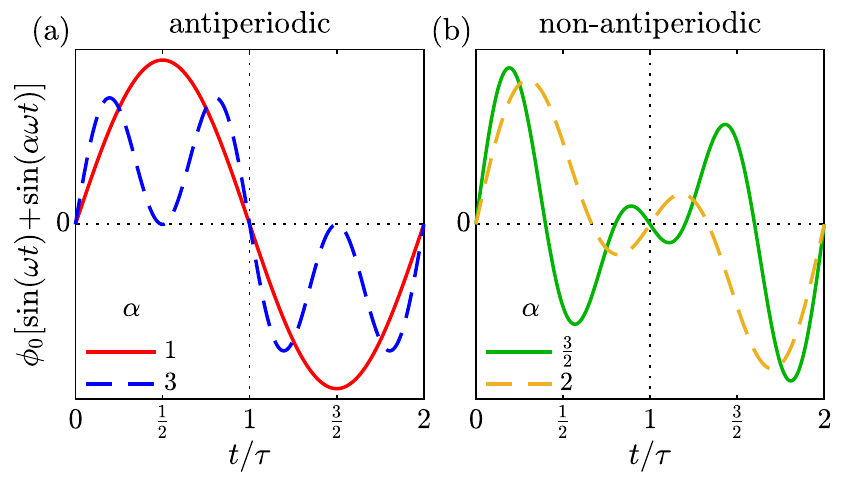}
\caption{Examples of antiperiodic (a), and non-antiperiodic (b), two-mode applied potentials $\phi_0\left[\sin(\omega t)+\sin(\alpha\omega t)\right]$.}
\label{Fig:rev_vs_irrev_pot}
\end{center}
\end{figure}

\autoref{Fig:rev_vs_irrev_pot} illustrates several examples of the antiperiodic and non-antiperiodic two-mode potentials. One can show that $\psi(t)$ is antiperiodic if $\alpha$, in its simplified fractional form, can be expressed as \{odd integer\}/\{odd integer\} (e.g., $\alpha=1,\;\tfrac{5}{3},\;3,\;5,\dots$). Otherwise, the two-mode potential is non-antiperiodic (e.g., $\alpha=2,\;\tfrac{4}{3},\;\tfrac{3}{2},\;4,\dots$). (See Hashemi et al. \cite{Aref2022PRE} for a simple proof.) Our numerical results for a wide range of $\alpha$ values corroborate our theory. For all antiperiodic potentials tested, the AREF is zero at the midplane, and is antisymmetric in space (e.g., $\alpha=1$ and $3$ in \autoref{Fig:AREF_two_mode_micronScale}(a)). Furthermore, a dissymmetric AREF with a nonzero value at the midplane is induced for non-antiperiodic potentials (e.g., $\alpha=2$ in \autoref{Fig:AREF_two_mode_micronScale}(a)). It should be noted though that the degree by which the AREF becomes dissymmetric is a complicated function of $\alpha$. However, regardless of the system parameters, $\alpha=2$ appears to induce the most significant dissymmetric behavior.
	
\begin{figure}[t]
\begin{center}
\centering
\setlength{\belowcaptionskip}{-20pt}
\includegraphics{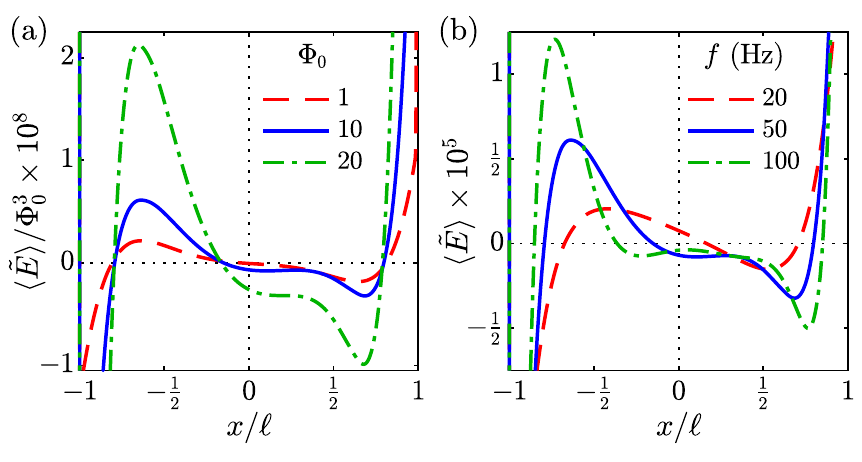}
\caption{Effects of the two-mode potential amplitude (a), and frequency (b), on the dissymmetric AREF. For visualization purposes the data in (a) are scaled by $\Phi_0^3$ with $\Phi_0=\phi_0/\phi_T$. Parameters: $\phi_0=10\phi_T$ (b), $f=\omega/(2\pi)=50$ $\mathrm{Hz}$ (a), $\alpha=2$, $2\ell=20$ $\mu\mathrm{m}$, $D_+=10^{-9}$ $\mathrm{m^2/s}$, $D_-/D_+=2$, $c^{\infty}=1$ $\mathrm{mM}$.}
\label{Fig:VolFreqEffect}
\end{center}
\end{figure}

In \autoref{Fig:VolFreqEffect}(a), we show the effect of the two-mode potential amplitude, $\phi_0$, on the induced AREF for $\alpha=2$. As a high-order nonlinear phenomena (cf. \autoref{sec:perturbation}), the dissymmetry rapidly grows with the amplitude. At sufficiently low amplitudes, the dissymmetry disappears and the AREF is almost antisymmetric (cf. \autoref{Fig:VolFreqEffect}(a), dashed red curve). An interesting finding here is that unlike the single-mode AREF, the curves of different voltage amplitudes do not collapse. There is no scaling factor (as a function of $\Phi_0$) that maps all of the AREF curves onto a master curve. This is in particular significant to the application of the AREF in particle height bifurcation \cite{Woehl2015,Bukosky2015}. It has been established that for a single mode potential, the AREF induced levitation height of charged colloids is insensitive to the amplitude of the potential, as are the zeros of the AREF, which determine approximately the heights at which the total force on a colloid is zero \cite{Aref2018,Bukosky2019}. Here, however, the zeros of the AREF, and hence, the levitation heights of the colloids, depend on the applied potential. This adds another parameter, along with the applied frequency, to tune the levitation height.

The effect of the applied frequency $f=\omega/(2\pi)$ is more complicated. Even for a single mode potential, it has been established that the spatial oscillation of the AREF (its shape) is very sensitive to frequency \cite{Aref2018,Aref2019}. For the two-mode potential, similar to the antisymmetric AREF, increasing the frequency amplifies the AREF peak magnitude in the bulk and shifts the peak location toward the electrodes \cite{Aref2018,Aref2019}, albeit through a more complex pattern (cf. \autoref{Fig:VolFreqEffect}(b)). Furthermore, we note that the dissymmetry intensifies substantially with frequency. More importantly, the sign of AREF at the midplane is changed upon changing the frequency.
		
Following Hashemi et al. \cite{Aref2018}, we have performed several consistency checks on the numerical results, such as the feasibility of the calculated instantaneous ion concentrations, electric field, and induced zeta potential at the electrode surface. Furthermore, the numerical solution converges and the total mass is conserved. We have inspected the total charge neutrality by $\int_{-\ell}^{\ell}\partial^2\langle\phi\rangle/{\partial x}^2dx=0$ and, alternatively, by $\langle E\rangle_{-\ell}=\langle E\rangle_{\ell}$. The condition $\int_{-\ell}^{\ell}\langle E\rangle=0$ is also checked to ensure that the numerical solution satisfies the boundary conditions. A concern in dynamic solution of the PNP equations under oscillatory polarization is that if the quasi-steady state conditions (harmonic solution) is achieved. We have accurately examined the present numerical results regarding the harmonic behavior.

We emphasize that our theory is not limited to any specific potential wave form. A general zero-mean function $\psi(t)$ with a period $2\tau$ has a Fourier series of the form $\psi(t)=\sum_{n=1}^{\infty}(a_n\cos(n\pi t/\tau)+b_n\sin(n\pi t/\tau))$, which is antiperiodic if $a_n=b_n=0$ for even $n$. Therefore, any antiperiodic $\psi(t)$ can be expressed as
\begin{equation}
  \psi(t)=\sum_{n=1,3,\dots}^{\infty}\left(a_n\cos\left(\frac{n\pi t}{\tau}\right)+b_n\sin\left(\frac{n\pi t}{\tau}\right)\right),
\end{equation}
and the ratio of any two frequencies will be the ratio of two odd integers.

\begin{figure}[t]
\begin{center}
\centering
\setlength{\belowcaptionskip}{-20pt}
\includegraphics{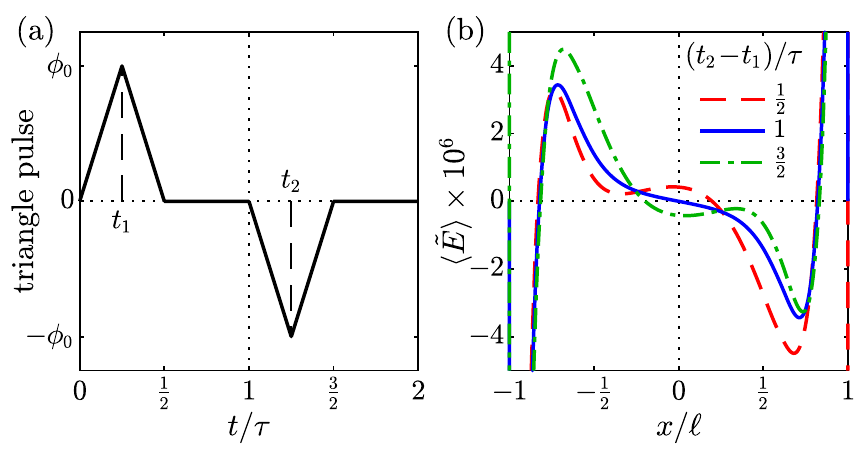}
\caption{Dissymmetric AREF due to a zero-time-average triangular pulse potential applied at $x=-\ell$. (a) Positive (at $t=t_1$) and negative (at $t=t_2$) triangular pulses of width $\tfrac{1}{2}\tau$ and amplitude $\phi_0$. (b) The corresponding induced AREF for different values of $(t_2-t_1)/\tau$. Parameters: $\phi_0=20\phi_T$, $1/(2\tau)=50$ $\mathrm{Hz}$, $2\ell=20$ $\mu\mathrm{m}$, $D_+=10^{-9}$ $\mathrm{m^2/s}$, $D_-/D_+=2$, $c^{\infty}=1$ $\mathrm{mM}$.}
\label{Fig:triangle_pot}
\end{center}
\end{figure}

It should be understood that for a given applied potential $\psi(t)$ in the first half of the period $t\in[0,\tau]$, there is a unique antiperiodic potential that occurs by setting $\psi(t+\tau)=-\psi(t)$. But an infinite number of non-antiperiodic potentials can be constructed. We demonstrate this argument for a triangular pulse of period $2\tau$, illustrated in \autoref{Fig:triangle_pot}. Two pulses of amplitude $\phi_0$ and width $\tfrac{1}{2}\tau$ are applied at $t_1=\tfrac{1}{4}\tau$ and $\tfrac{3}{4}\tau\le t_2\le\tfrac{7}{4}\tau$. We keep $t_1$ fixed and vary $t_2$ to cover all possible cases. The induced AREFs are shown in \autoref{Fig:triangle_pot}(b) for different $t_2-t_1$ values. The AREF is antisymmetric only if $t_2-t_1=\tau$ for which the potential in the second half period becomes the negative of that in the first half (\autoref{Fig:triangle_pot}(b), solid blue curve). All other constructions yield a dissymmetric AREF. It is interesting to note that the cases $t_2-t_1=\tfrac{1}{2}\tau$ (consecutive pulses in the first half) and $t_2-t_1=\tfrac{3}{2}\tau$ (maximally apart pulses) provide the maximum dissymmetry and are mirrored. A simple time shift $t\to\tfrac{3}{2}\tau$ shows that the condition of maximally apart pulses is actually the negative version of the back-to-back pulses, and therefore yields the mirrored version of the AREF (cf. \autoref{Fig:AREF_reversePotential}(b)).
	
\section{Perturbation analysis}\label{sec:perturbation}
In this section, we develop a perturbation expansion solution (in terms of $\Phi_0=\phi_0/\phi_T$) to the PNP equations for a two-mode applied potential. Our results indicate that the dissymmetric AREF is a consequence of the odd-order ($3,\;5,\dots$) perturbation terms. Indeed, an extension to the second-order solution derived by Hashemi et al. \cite{Aref2020SM} predicts that, regardless of $\alpha$, the induced AREF under two-mode polarization is simply the superposition of the AREF due to each mode ($\omega$ and $\alpha\omega$) and is, therefore, antisymmetric. The third-order solution provides some insights, indicating a dissymmetric AREF for $\alpha=2$. However, the third-order solution is zero for all other $\alpha$ values, including the ones for which we expect a dissymmetric behavior. We argue that the numerically obtained dissymmetric AREF for other $\alpha$ values is a contribution of yet higher odd-order solutions (e.g., $5,\;7,\dots$).

\subsection{Second-order perturbation solution (single-mode)}
First, we briefly review the second-order solution to the PNP equations derived by Hashemi et al. \cite{Aref2020SM} for a single-mode sinusoidal potential. The dimensionless PNP equations can be expressed as 
\begin{align}
-\frac{\partial^2\tilde{\phi}}{{\partial\tilde{x}}^2}&=\tfrac{1}{2}\tilde{n}_+-\tfrac{1}{2}\tilde{n}_-,\label{Eq:RA_PDL}\\
\frac{\partial \tilde{n}_{\pm}}{\partial\tilde{t}}&=-\frac{\tilde{j}_{\pm}}{\partial\tilde{x}},\label{Eq:RA_NPDL}
\end{align}
where the ion flux $\tilde{j}_{\pm}$ is given by
\begin{equation}
\tilde{j}_{\pm}=-\frac{1}{1\mp\beta}\left[\frac{\partial \tilde{n}_{\pm}}{\partial\tilde{x}}\pm\left(\tilde{n}_{\pm}\frac{\partial\tilde{\phi}}{\partial\tilde{x}}\right)\right]\!\!,
\end{equation}
and the initial and boundary conditions are
\begin{align}
\tilde{n}_{\pm}(\tilde{x},0)&=1,\label{Eq:RA_ICDL}\\
\tilde{\phi}(\pm\kappa\ell,\tilde{t})&=\mp\tilde{\psi}(\tilde{t}),\label{Eq:RA_phiBCDL}\\
\tilde{j}_{\pm}(\pm\kappa\ell,\tilde{t})&=0, \label{Eq:RA_nBCDL}
\end{align}
with $\tilde{\psi}(\tilde{t})=\Phi_0\sin(\Omega\tilde{t})$.

Here, the dimensionless variables and parameters are defined as
\begin{alignat}{2}
  \tilde{n}_{\pm}\!=\!\frac{n_{\pm}}{n^{\infty}},&\quad\tilde{\phi}\!=\!\frac{\phi}{\phi_T},\quad\tilde{x}\!=\!\kappa x,\quad\tilde{t}\!=\!t\kappa^2D,&&\\
  \Phi_0\!=\!\frac{\phi_0}{\phi_T},&\quad\Omega\!=\!\frac{\omega}{\kappa^2D},\quad\kappa\ell,\quad\beta\!=\!\frac{D_+\!-\!D_-}{D_+\!+\!D_-},&&
\end{alignat}
where
\begin{equation}
D=\frac{2D_+D_-}{D_++D_-},\quad\kappa^{-1}=\sqrt{\frac{\varepsilon k_BT}{2e^2n^{\infty}}}.
\end{equation}

In the limit of small potentials ($\Phi_0\ll1$), one can approximate the solution by the power series
\begin{align}
  \tilde{n}_{\pm}(\tilde{x},\tilde{t})\!&=\!\tilde{n}_{\pm}^{(0)}\!+\!\Phi_0\tilde{n}_{\pm}^{(1)}\!+\!\Phi_0^{2}\tilde{n}_{\pm}^{(2)}\!+\!\Phi_0^{3}\tilde{n}_{\pm}^{(3)}\!+\!\dots\\
  \tilde{\phi}(\tilde{x},\tilde{t})\!&=\!\tilde{\phi}^{(0)}\!+\!\Phi_0\tilde{\phi}^{(1)}\!+\!\Phi_0^{2}\tilde{\phi}^{(2)}\!+\!\Phi_0^{3}\tilde{\phi}^{(3)}\!+\!\dots
\end{align}
where the superscript $(i)$ denotes the perturbation order. We substitute the above series for $\tilde{n}_{\pm}$ and $\tilde{\phi}$ into the PNP equations (Eqs.~\ref{Eq:RA_PDL} and \ref{Eq:RA_NPDL}), and the corresponding initial and boundary conditions (Eqs.~\ref{Eq:RA_ICDL}--\ref{Eq:RA_nBCDL}), equate terms with common powers of $\Phi_0$, and then solve separately for the zeroth, first, and second order solutions.

\subsubsection{Zeroth and first order solutions}
The zeroth-order solution is simply $\tilde{n}_{\pm}^{(0)}=1,\;\tilde{\phi}^{(0)}=0$, and the first-order solution is given by
\begin{align}
  \tilde{n}_{\pm}^{(1)}(\tilde{x},\tilde{t})\!&=\!\mathrm{Im}\!\left[\hat{n}_{\pm}^{(1)}e^{i\Omega\tilde{t}}\right]\!\!,\\
  \tilde{\phi}^{(1)}(\tilde{x},\tilde{t})\!&=\!\mathrm{Im}\!\left[\hat{\phi}^{(1)}e^{i\Omega\tilde{t}}\right]\!\!,
\end{align}
where $\hat{n}_{\pm}^{(1)}$ and $\hat{\phi}^{(1)}$ are odd (with respect to $x$) complex functions given by Eqs.~(30)--(37) in Hashemi et al. \cite{Aref2020SM}.
	
\subsubsection{Second-order solution}
The second-order governing equations for $\tilde{n}_{\pm}^{(2)}$ and $\tilde{\phi}^{(2)}$ are \cite{Aref2020SM}:
\begin{equation}
  \frac{\partial\tilde{n}_{\pm}^{(2)}}{\partial\tilde{t}}+\frac{\partial\tilde{j}_{\pm}^{\,(2)}}{\partial\tilde{x}}=0,
  \label{Eq:RA_NP2}
\end{equation}
\begin{equation}
\frac{\partial^2\tilde{\phi}^{(2)}}{{\partial\tilde{x}}^2}+\tfrac{1}{2}\tilde{n}_+^{(2)}-\tfrac{1}{2}\tilde{n}_-^{(2)}=0.
\label{Eq:RA_P2}
\end{equation}
Here the ion flux is given by
\begin{equation}
\tilde{j}_{\pm}^{\,(2)}(\tilde{x},\tilde{t})=-\frac{1}{1\mp\beta}\left[\frac{\partial\tilde{n}_{\pm}^{(2)}}{\partial\tilde{x}}\!\pm\!\frac{\partial\tilde{\phi}^{(2)}}{\partial\tilde{x}}\!+\!s_{\pm}^{(2)}\right]\!\!,
\end{equation}
with
\begin{equation}
s_{\pm}^{(2)}=\pm\tilde{n}_{\pm}^{(1)}\frac{\partial\tilde{\phi}^{(1)}}{\partial\tilde{x}}=\mp\tilde{n}_{\pm}^{(1)}\tilde{E}^{(1)}.\label{Eq:RA_source}
\end{equation}
The boundary conditions are
\begin{equation}
\tilde{\phi}^{(2)}(\pm\kappa\ell,\tilde{t})=0,\quad\tilde{j}_{\pm}^{\,(2)}(\pm\kappa\ell,\tilde{t})=0.
\end{equation}
	
Hashemi et al. focused only on the second-order time-average electric field which was shown to accurately predict AREF via a simple linear ordinary differential equation (ODE), the AREF equation \cite{Aref2020SM}:
\begin{equation}
\frac{\partial^2\langle\tilde{E}^{(2)}\rangle}{{\partial\tilde{x}}^2}-\langle\tilde{E}^{(2)}\rangle=f^{(2)},
\label{Eq:RA_AREF_2nd}	
\end{equation}
with
\begin{equation}
f^{(2)}=\tfrac{1}{2}\langle s_-^{(2)}\rangle-\tfrac{1}{2}\langle s_+^{(2)}\rangle.
\end{equation}
	
An important point is that Hashemi et al. \cite{Aref2020SM} assumed $\langle\tilde{E}^{(2)}\rangle=0$ at the boundaries $\tilde{x}=\pm\kappa\ell$, which was consistent with all of the numerical results for single-mode sinusoidal applied potentials. Here we justify this assumption. One can easily show that the RHS $f^{(2)}$ is an odd function. Recall that $\tilde{n}_{\pm}^{(1)}$ and $\tilde{\phi}^{(1)}$ are odd which makes $\tilde{E}^{(1)}=-\partial\tilde{\phi}^{(1)}/\partial\tilde{x}$ an even function. Hence, $f^{(2)}$ which includes $\mp\tilde{n}_{\pm}^{(1)}\tilde{E}^{(1)}$ is odd, and yields an odd second-order solution $\langle\tilde{E}^{(2)}\rangle$ (antisymmetric). Additionally, total charge neutrality requires $\langle\tilde{E}^{(2)}\rangle_{\kappa\ell}=\langle\tilde{E}^{(2)}\rangle_{-\kappa\ell}=K$. But for $\langle\tilde{E}^{(2)}\rangle$ to be odd $K$ has to be $0$ which justifies the assumption. An alternative would be to solve for the potential via the following rank-deficient problem:
\begin{equation}
  -\frac{\partial^3\langle\tilde{\phi}^{(2)}\rangle}{{\partial\tilde{x}}^3}+\frac{\langle\partial\tilde{\phi}^{(2)}\rangle}{\partial\tilde{x}}=f^{(2)},
  \label{Eq:RA_pot-ave_2nd}
\end{equation}
subject to $\langle\tilde{\phi}^{(2)}\rangle=0$ at $\tilde{x}=\pm\kappa\ell$. Solving the problem using singular-value-decomposition (SVD) along with the total charge neutrality condition,
\begin{equation}
  \int_{-\kappa\ell}^{\kappa\ell}\frac{\partial^2\langle\tilde{\phi}^{(2)}\rangle}{{\partial\tilde{x}}^2}d\tilde{x}=0,
\end{equation}
yields the same results as that of setting $\langle\tilde{E}^{(2)}\rangle=0$ at $\tilde{x}=\pm\kappa\ell$.
	
\subsection{Second-order perturbation solution (two-mode)}
In this section we follow the same procedure to derive the perturbation solution for the two-mode potential $\tilde{\psi}(\tilde{t})=\Phi_0\left[\sin(\Omega\tilde{t})+\sin(\alpha\Omega\tilde{t})\right]$.
\subsubsection{Zeroth and first order solutions}
For the two-mode system, the zeroth-order solution remains unchanged. The first-order solution is obtained by superposition of the first-order solutions due to $\Omega$ and $\alpha\Omega$ modes:
\begin{align}
  \tilde{n}_{\pm}^{(1)}(\tilde{x},\tilde{t})\!&=\!\mathrm{Im}\!\left[\hat{n}_{\pm}^{(1)}\sO e^{i\Omega\tilde{t}}\!+\!\hat{n}_{\pm}^{(1)}\saO e^{i\alpha\Omega\tilde{t}}\right]\!\!,\\
  \tilde{\phi}^{(1)}(\tilde{x},\tilde{t})\!&=\!\mathrm{Im}\!\left[\hat{\phi}^{(1)}\sO e^{i\Omega\tilde{t}}\!+\!\hat{\phi}^{(1)}\saO e^{i\alpha\Omega\tilde{t}}\right]\!\!.
\end{align}
	
\subsubsection{Second-order solution}
\begin{figure}[t]
  \begin{center}
    \centering
    \setlength{\belowcaptionskip}{-20pt}
    \includegraphics{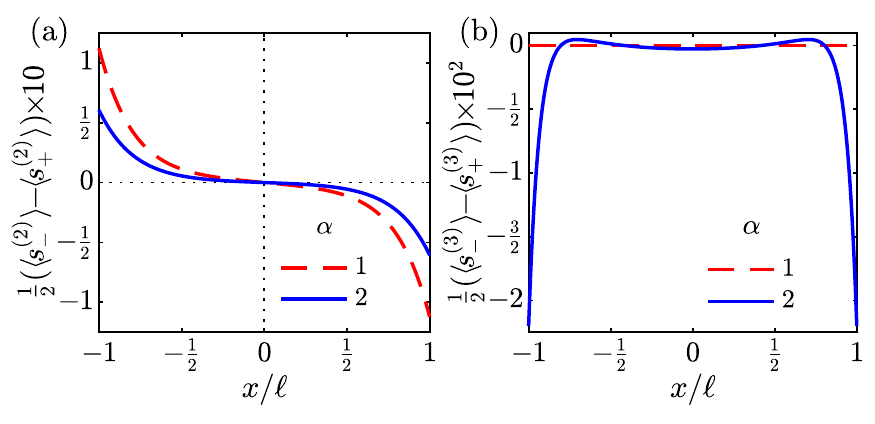}
    \caption{RHSs of the second-order (a), and third-order (b), AREF equations for $\alpha=1$ and $2$. Parameters: $\Omega=0.1$, $\beta=-1/3$, $\kappa\ell=5$.}
    \label{Fig:RA_Fig_source}
  \end{center}
\end{figure}

One can show that $f^{(2)}$ for the two-mode case is the superposition of the RHSs due to $\Omega$ and $\alpha\Omega$. Hence, following Hashemi et al. \cite{Aref2020SM}, we can solve the problem for the modes $\Omega$ and $\alpha\Omega$ separately without any extra complications. More importantly, it shows that the second-order solution does not provide any information regarding the possible dissymmetric behavior of AREF. As mentioned, the RHS of the second-order AREF equation is odd and yields antisymmetric contributions (cf. \autoref{Fig:RA_Fig_source}(a)).
	
Consequently, we suspect that the dissymmetric AREF is a third-order phenomena. To investigate this hypothesis we need to first fully solve the second-order system to find the temporal and spatial distributions of the ion concentrations and the potential/electric field. The first and second order solutions will provide the RHS of the third-order AREF equation.
	
We use Fourier series to solve the problem. Consider solutions of the form:
\begin{align}
  \tilde{n}_{\pm}^{(2)}&=\sum_{k=-\infty}^{\infty}c_k^{\pm}e^{ik\Omega_c\tilde{t}},\label{Eq:RA_FS_NP2}\\
  \tilde{\phi}^{(2)}&=\sum_{k=-\infty}^{\infty}c_k^{\phi}e^{ik\Omega_c\tilde{t}}.\label{Eq:RA_FS_P2}
\end{align}
Here $\Omega_c=\Omega\gcd(1,\alpha)$ is the common angular frequency of the applied potential, i.e., $2\pi/\Omega_c$ is the period of the two-mode applied potential/harmonic solution. Inserting \autoref{Eq:RA_FS_NP2} and \autoref{Eq:RA_FS_P2} into the second-order governing equations and boundary conditions yields the following differential equations for the coefficients of the Fourier series.
	
\begin{enumerate}[wide, labelwidth=!, labelindent=0pt,label=(\roman*)]
\item{$k=0$}
\begin{subequations}
\begin{gather}
  \frac{\partial^2c_0^{\pm}}{{\partial\tilde{x}}^2}\pm\frac{\partial^2c_0^{\phi}}{{\partial\tilde{x}}^2}=-\bigg\langle\!\frac{\partial s_{\pm}^{(2)}}{\partial\tilde{x}}\!\bigg\rangle,\label{Eq:RA_NP_c0}\\
  \frac{\partial^2c_0^{\phi}}{{\partial\tilde{x}}^2}+\tfrac{1}{2}c_0^+-\tfrac{1}{2}c_0^-=0,\label{Eq:RA_P_c0}\\
  \frac{\partial c_0^{\pm}}{\partial\tilde{x}}\pm\frac{\partial c_0^{\phi}}{\partial\tilde{x}}=-\langle s_{\pm}^{(2)}\rangle\quad\text{at}\quad \tilde{x}=\pm\kappa\ell,\label{Eq:RA_NP_BC_c0}\\
  c_0^{\phi}(\pm\kappa\ell)=0.\label{Eq:RA_P_BC_c0}
\end{gather}
\end{subequations}

\autoref{Eq:RA_NP_c0} and \autoref{Eq:RA_NP_BC_c0} can be combined to yield
\begin{equation*}
  \frac{\partial c_0^{\pm}}{\partial\tilde{x}}\pm\frac{\partial c_0^{\phi}}{\partial\tilde{x}}+\langle s_{\pm}^{(2)}\rangle=0.
\end{equation*}
Now, taking the derivative of \autoref{Eq:RA_P_c0} with respect to $\tilde{x}$, and subsequent substitution of $\partial c_0^{\pm}/\partial\tilde{x}$ from the above equation results in
\begin{equation}
  \frac{\partial^3c_0^{\phi}}{{\partial\tilde{x}}^3}-\frac{\partial c_0^{\phi}}{\partial\tilde{x}}=\tfrac{1}{2}\langle s_+^{(2)}\rangle-\tfrac{1}{2}\langle s_-^{(2)}\rangle.
  \label{Eq:RA_AREF_2nd_c0_phi}
\end{equation}
Note that we could write this equation in terms of $c_0^{E}=-\partial c_0^{\phi}/\partial\tilde{x}$ to get:
\begin{equation*}
  \frac{\partial^2c_0^{E}}{{\partial\tilde{x}}^2}-c_0^{E}=\tfrac{1}{2}\langle s_-^{(2)}\rangle-\tfrac{1}{2}\langle s_+^{(2)}\rangle,
\label{Eq:RA_AREF_2nd_c0_E}
\end{equation*}
which is equivalent to the linear ODE used by Hashemi et al. \cite{Aref2020SM} to find AREF (\autoref{Eq:RA_AREF_2nd}).
	
\item{$k=-\infty,\dots,\infty$}
\begin{subequations}
  \begin{gather}
    \!\!\!\frac{\partial^2c_k^{\pm}}{{\partial\tilde{x}}^2}\pm\frac{\partial^2c_k^{\phi}}{{\partial\tilde{x}}^2}\!-\!ik\Omega_c(1\!\mp\!\beta)c_k^{\pm}\!=\!-\bigg\langle\!\!\frac{\partial s_{\pm}^{(2)}}{\partial\tilde{x}}e^{-ik\Omega_c\tilde{t}}\!\bigg\rangle\!,\!\!\!\label{Eq:RA_NP_ck}\\
    \frac{\partial^2c_k^{\phi}}{{\partial\tilde{x}}^2}+\tfrac{1}{2}c_k^+-\tfrac{1}{2}c_k^-=0,\label{Eq:RA_P_ck}\\
    \frac{\partial c_0^{\pm}}{\partial\tilde{x}}\pm\frac{\partial c_0^{\phi}}{\partial\tilde{x}}=-\Big\langle s_{\pm}^{(2)}e^{-ik\Omega_c\tilde{t}}\Big\rangle\quad\text{at}\quad \tilde{x}=\pm\kappa\ell,\label{Eq:RA_NP_BC_ck}\\
    c_k^{\phi}(\pm\kappa\ell)=0.\label{Eq:RA_P_BC_ck}
  \end{gather}
\end{subequations}

\end{enumerate}

We solve the above equations numerically to find $c_k^{\pm}$ and $c_k^{\phi}$ ($k=-\infty,\dots,\infty$). Of course we do not need to solve for $k\to\pm\infty$ and we truncate the series at a much smaller $k$ value (i.e., $k=-K,\dots,K$). Considering the source terms (\autoref{Eq:RA_source}) as a multiplication of first-order solutions, the maximum possible frequency component can be obtained. For example, for a single-mode applied potential of angular frequency $\Omega$, the first-order terms (i.e., $\tilde{n}_{\pm}^{(1)},\;\tilde{E}^{(1)}$) oscillate with $\Omega$ and hence, the highest possible angular frequency of the source terms becomes $2\Omega=2\Omega_c$ and $K=2$. Similarly, for a two-mode system ($\Omega$ and $\alpha\Omega$), the highest angular frequency component, given $\alpha\ge1$, is $2\alpha\Omega=2\alpha\Omega_c/\gcd(1,\alpha)\Rightarrow K=2\alpha/\gcd(1,\alpha)$. Note that when $\alpha=1$, $K=2$ is recovered.
	
\subsubsection{Third-order perturbation}\label{Sec:RA_third-order}
The third-order system of equations is identical to that of the second order, except the source terms which are
\begin{equation}
  s_{\pm}^{(3)}=\pm\left(\tilde{n}_{\pm}^{(1)}\frac{\partial\tilde{\phi}^{(2)}}{\partial\tilde{x}}\!+\!\tilde{n}_{\pm}^{(2)}\frac{\partial\tilde{\phi}^{(1)}}{\partial\tilde{x}}\right)\!\!.
\end{equation}
The third-order AREF equation is given by
\begin{equation}
  \frac{\partial^2\langle\tilde{E}^{(3)}\rangle}{{\partial\tilde{x}}^2}-\langle\tilde{E}^{(3)}\rangle=f^{(3)},
\label{Eq:RA_AREF_3rd}	
\end{equation}
where
\begin{equation}
f^{(3)}=\tfrac{1}{2}\langle s_-^{(3)}\rangle-\tfrac{1}{2}\langle s_+^{(3)}\rangle.
\end{equation}
	
An analysis of the RHS $f^{(3)}$ provides some insight regarding the dissymmetric AREF. We note that the second-order solutions $\tilde{\phi}^{(2)}$ and $\tilde{n}_{\pm}^{2}$ are even functions. Therefore, $f^{(3)}$ is an even function in space which, in turn, indicates that $\tilde{E}^{(3)}$ is even. Hence, when $f^{(3)}\neq0$, the third-order electric field adds a dissymmetric contribution to the overall AREF. \autoref{Fig:RA_Fig_source}(b) shows the calculated $f^{(3)}$ for $\alpha=1$ and $2$. We note that the $f^{(3)}\neq0$ when $\alpha=2$, corroborating the observed dissymmetric behavior. However, $f^{(3)}=0$ for all other $\alpha$ values, including the ones that are expected (based on the numerical results) to result in dissymmetry. We argue that higher odd-order solutions ($5,\;7,\dots$) are responsible for the observed dissymmetry for other $\alpha$ values such as $\tfrac{3}{2},\;\tfrac{4}{3},\;4$.
	
\subsubsection{Generalization}
The $k^{\text{th}}$ order solution ($k\ge2$) contributes to the overall AREF via the ODE
\begin{equation}
\frac{\partial^2\langle\tilde{E}^{(k)}\rangle}{{\partial\tilde{x}}^2}-\langle\tilde{E}^{(k)}\rangle=f^{(k)}.
\end{equation}
It can be shown that the source terms are odd functions with respect to the midplane for even $k$ values, yielding an antisymmetric contributions to the solution. The source terms for odd orders are even functions in space, resulting in even contributions to the electric field. Therefore, if $s^{(k)}$ is nonzero for $k=3,\;5,\dots$, the perturbation solution suggests a dissymmetric shape for the AREF.

\section{\uppercase{Conclusions}}
In summary, our results show that ions and charged colloids can be concentrated to one side of a slit channel, or another, by tuning the applied potential waveform. We demonstrate that the induced AREF between parallel electrodes by a non-antiperiodic electric potential is spatially dissymmetric. An intriguing implication is then that swapping the powered and grounded electrodes of an electrochemical cell alters the system behavior, an observation at odds with the classical understanding of the electrokinetics. The dissymmetric AREF can tremendously change the design of electrokinetic systems and their applications. It was recently shown at length that the AREF-induced electrophoretic forces are several orders of magnitude larger that gravitational and colloidal forces \cite{Aref2018,Bukosky2019,Aref2019,Aref2020}. Researchers can therefore use the dissymmetric AREF to design electrochemical cells that selectively (to some extent) separate charged colloidal particles or bioparticles near the powered or the grounded electrodes. Moreover, the sole physical implications of the dissymmetric AREF opens a new chapter for the researchers in the electrokinetic community.

\vspace{1cm}\noindent\textbf{Acknowledgments.} This material is based upon work partially supported by the National Science Foundation under Grants No. DMS-1664679 and No. CBET-2125806.


%

\end{document}